\documentstyle[multicol,aps]{revtex}

\begin{document}
\title{Deterministic Soluble Model of Coarsening}
\author{L.~Frachebourg and P.~L.~Krapivsky} 
\address{Center for Polymer Studies and Department of Physics,
Boston University, Boston, MA 02215}
\maketitle
\begin{abstract}

We investigate a 3-phase deterministic one-dimensional phase
ordering model in which interfaces move ballistically and annihilate
upon colliding.  We determine analytically the autocorrelation
function $A(t)$.  This is done by computing generalized first-passage
type probabilities $P_n(t)$ which measure the fraction of space
crossed by exactly $n$ interfaces during the time interval $(0,t)$, and then
expressing the autocorrelation function via $P_n$'s.  We further reveal
the spatial structure of the system by analyzing the domain size 
distribution.

\end{abstract}
\begin{multicols}{2}

\section{INTRODUCTION AND THE MODEL}

We examine phase ordering dynamics in a one-dimensional system with
three equilibrium states.  In our model, interfaces between dissimilar
domains undergo ballistic motion and annihilate upon colliding.  The
process is thus deterministic although randomness is hidden in the
initial conditions.  Given an appealing simplicity of the rules
governing the dynamics, it is not surprising that this process, and
its generalizations, naturally arise in different contexts ranging
from ballistic annihilation\cite{els,brl,krl,Piasecki,Droz} to growth
processes\cite{Krug,sekimoto,ben,gold} and dynamics of interacting
populations\cite{bramson,fisch,lpe}.  Different viewpoints on the same
model are very useful in that they suggest investigation of several
correlation functions, some of them may be clearly interesting from one
point of view while could hardly be thought from the other point of
view.  One such correlation function, namely the autocorrelation
function to be determined below, naturally appears in the context of
population dynamics\cite{lpe}; from other viewpoints, {\it e.g.} in the
original framework of ballistic annihilation\cite{els} it is not clear
how to define the autocorrelation function.

We start by describing the two-velocities ballistic annihilation model 
and reminding its known basic properties\cite{els,Piasecki}.  
The model assumes that interfaces may
have two different velocities, $\pm 1$ without loss of generality, and
the densities of both populations of interfaces are equal each other
(otherwise the minority population quickly disappears). The interfaces
are initially randomly distributed according to a Poisson distribution.
The model exhibits a two-length spatial structure, with length scale
$\ell(t)\sim \sqrt{t}$ describing the average distance between
neighboring interfaces moving in the same direction, and the length scale
${\cal L}(t)\sim t$ describing the typical distance between
neighboring interfaces moving in the opposite directions.  As we shall
see below, however, the growth law for $\ell(t)$ cannot fully
characterize the spatial structure -- other natural measures of the
spacing between similar neighboring interfaces behave differently, 
{\it e.g.}, the rms separation $\ell_2(t)=\sqrt{\langle x^2\rangle}$ 
grows as $t^{3/4}$.  We shall 
argue below that all these length scales can be understood as the 
outcome of the competition between the length scale ${\cal O}(1)$
characterizing initial data and the ballistic length scale 
${\cal L}(t)\sim t$.

On the language of phase ordering dynamics, the two-velocity ballistic
annihilation model may be treated as the 3-phase, or 3-color, process
with deterministic nonconservative dynamics.  Indeed, imagine that the
one-dimensional line is drawn in three colors, say red, green, and
blue.  Let the interface between red and green domains always moves
inside the green one, the interface between green and blue domains
moves inside the blue one, and the interface between blue and red
domain moves inside the red one.  Then the autocorrelation function
$A(t)$ is defined as the probability that at a given point and 
at time $t$ the color is identical to the initial color.
In the dynamics of interacting populations, this model mimics a
3-species cyclic food chain\cite{lpe}.

The rest of this paper is organized as follows.  Generalized 
first-passage probabilities are determined in section II.
Section III contains a calculation of the autocorrelation function. 
The domain size distribution is analyzed in section IV. The last
section V provides a summary and an outlook.

\section{GENERALIZED FIRST-PASSAGE PROBABILITIES}

Our first goal is to compute $P_n(t)$ which measures the fraction of
space crossed by exactly $n$ interfaces during the time interval
$(0,t)$.  Equivalently, $P_n(t)$ is the probability that a point has
undergone exactly $n$ changes of color.  Clearly, the color of
arbitrary point changes cyclically with period 3, so the
autocorrelation function is found from relation

\begin{equation}    
A(t)=\sum_{n=0}^\infty P_{3n}(t).
\label{au}
\end{equation}

To determine $P_n(t)$, it proves convenient to consider an auxiliary
one-sided problem with a finite number of interfaces on one side of a
target point.  Namely, imagine that we have $N$ interfaces to the
right of the origin (the target point).  What is the probability
$Q_n(N)$ that exactly $n$ interfaces will cross the origin?  To solve
for $Q_n(N)$, we construct the following discrete random walk: Let
$S_0=0$ and $S_i$ are defined recursively via
$S_i=S_{i-1}+v_i,\,i=1,\ldots,N$, where $v_i=\pm 1$ is the velocity of
the $i^{\rm th}$ interface.  Thus we indeed have a random walk
$(i,S_i)$ starting from the origin, with $i$ being a time-like
variable and $S_i$ a displacement.  The crucial point is that the
number of interfaces which will cross the origin is given by the
absolute value of the minimum of the random walk.  Thus we identify
$Q_n(N)$ with probability that an $N$-step random walk starting at the
origin has a minimum at $-n$.  
This probability is simply found to be \cite{feller}
\begin{equation}    
Q_n(N)=\tilde Q_n(N)+\tilde Q_{n+1}(N),
\label{qnN}
\end{equation}
with
\begin{equation}    
\tilde Q_n(N)=
{1\over 2^N}\,{N!\over \left({N+n\over 2}\right)!
\left({N-n\over 2}\right)!}
\label{tqnN}
\end{equation}
if $n$ and $N$ have the same parity; otherwise, $\tilde Q_n(N)=0$.

Before returning to the original two-sided problem we consider the
one-sided problem with {\it infinite} number of interfaces initially
placed to the right of the origin at random with density one.  During
the time interval $(0,t)$ interfaces initially located at distances
$x\leq t$ could cross the origin.  Clearly, the probability $Q_n(t)$
that exactly $n$ interfaces cross the origin up to time $t$ is

\begin{equation}    
Q_n(t)=\sum_{N=n}^\infty Q_n(N)\,{t^N e^{-t}\over N!}.
\label{qnt}
\end{equation}
Substituting (\ref{qnN}) and (\ref{tqnN}) into (\ref{qnt}) yields
\begin{equation}    
Q_n(t)=e^{-t}\left[I_n(t)+I_{n+1}(t)\right],
\label{iqnt}
\end{equation}
where $I_n$ denotes the modified Bessel function of order $n$.
If the origin is not crossed by a right moving interface up to time $t$, 
an interface starting from the origin and moving with $+1$ velocity 
will survive up to time $t/2$.  Thus the surviving probability, 
$S(t)$, of an interface is given by 

\begin{equation} 
S(t)=Q_0(2t)=e^{-2t}\left[I_0(2t)+I_1(2t)\right].
\label{St}
\end{equation}

First-passage probabilities $P_n(t)$ corresponding to the two-sided
problem are readily expressed via one-sided probabilities $Q_n(t)$
after realizing that in a configuration with $n$ interfaces crossing
the origin in the right-sided version, and $k$ interfaces crossing the
origin in the left-sided version, the total crossing number in the
two-sided version is equal to $\max(k,n)$, as illustrated in Fig.~1.
Thus we arrive at the relationship

\begin{equation}    
P_n(t)=2Q_n(t)\sum_{k=0}^n Q_k(t)-Q_n(t)^2,
\label{pnt}
\end{equation}
with factor 2 accounting for the fact that smaller number $k$ of
crossing interfaces can come both from the left and right.  We have
subtracted the last quantity $Q_n(t)^2$ which has been counted twice
in the summation.  As a useful check of self-consistency we verify
that the normalization condition,
 
\begin{equation}    
\sum_{n=0}^\infty P_n(t)=1, 
\label{norm}
\end{equation}
is satisfied.  Indeed, Eq.~(\ref{pnt}) implies
$\sum P_n=(\sum Q_n)^2$, and the
latter sum is shown to be equal to one by using Eq.~(\ref{iqnt}) and
identity $I_0(t)+2\sum_{j\geq 1}I_j(t)=e^t$\cite{bender}.

Note that $P_n$'s, and especially the first ``persistence''
probability $P_0(t)$, have attracted a considerable recent interest,
see {\it e.g.}\cite{dbg,kbr,bdg,der,elp,stev}.  These quantities can
be thought as first-passage time probabilities in the interacting
particle systems\cite{lig}.  Given the importance of the first-passage
type quantities in the classical probability theory\cite{feller} one
can envision numerous applications of $P_n$'s in the interacting
particle systems.  However, apart from a few findings in the framework
of mean-field approach (more precisely, for interacting particle
systems on a complete graph)\cite{lpe,elp} and a limiting analytical
solution for the 1D voter model\cite{elp}, no exact results are
available.  The model we consider here is an exception in that the
complete analytical solution for $P_n$'s exists, see
(\ref{iqnt})--(\ref{pnt}).  In particular, we have 
$P_0(t)=Q_0^2(t)\simeq 2(\pi t)^{-1}$.   

To make results more transparent, it is useful to express solutions 
in the scaling limit
\begin{equation}    
n\to \infty, \quad
t\to \infty, \quad
z={n\over \sqrt{2t}}={\rm finite}.
\label{scal}
\end{equation}
Making use of the scaling behavior of the modified Bessel 
functions\cite{bender}, 
$I_n(t)\simeq (2\pi t)^{-1/2}\exp(t-n^2/2t)$, we find
\begin{equation}    
Q_n(t)\simeq 
\sqrt{2\over \pi t}\,e^{-z^2}
\label{1-scal}
\end{equation}
for the one-sided probabilities and
\begin{equation}
P_n(t)\simeq 
\sqrt{8\over \pi t}\,e^{-z^2}\,{\rm Erf}(z)
\label{2-scal}
\end{equation}
for the two-sided probabilities.

\section{THE AUTOCORRELATION FUNCTION}

The scaling expression of Eq.~(\ref{2-scal})
does not allow to obtain non trivial long-time behavior of the 
autocorrelation function.
Indeed, substituting (\ref{2-scal}) into (\ref{au}) yields 
$A(t)\simeq 1/3$.  We should therefore return to exact relations
(\ref{iqnt})--(\ref{pnt}).  We also extract the trivial $A(\infty)=1/3$
factor and consider three autocorrelation functions,

\begin{equation}
A_{\alpha}(t)=\sum_{n=0}^{\infty}P_{3n+\alpha}(t)-{1\over 3},
\label{auk}
\end{equation}
describing three possible color outcomes at time $t$, the same (say
red) color as initially corresponds to $\alpha=0$, $A_0(t)\equiv
A(t)-1/3$; the ``next'' blue color corresponds to $\alpha=1$; and
finally the green color corresponds to $\alpha=2$.

All three autocorrelation functions $A_{\alpha}(t)$ exhibit similar
asymptotic behavior; additionally, they are related by identity
$A_0(t)+A_1(t)+A_2(t)\equiv 0$.  Combining (\ref{auk})
and (\ref{norm}) we obtain

\begin{eqnarray}
3A_0(t)&=&3\sum_{n=0}^{\infty}P_{3n}(t)-1\nonumber\\
&=&\sum_{n=0}^{\infty}\left[(P_{3n}-P_{3n-1})+(P_{3n}-P_{3n+1})\right]
\label{au1}
\end{eqnarray}
where $P_{-1}\equiv 0$.
Eq.~(\ref{pnt}) allows us to express $P_n$'s via $Q_n$'s. 
Thus we get
\begin{eqnarray}
P_{3n}-P_{3n-1}&=&Q^2_{3n}+Q^2_{3n-1}\nonumber\\
&+&2(Q_{3n}-Q_{3n-1})\sum_{k=0}^{3n-1}Q_k
\label{pn1}
\end{eqnarray}
and
\begin{eqnarray}
P_{3n}-P_{3n+1}&=&Q^2_{3n}-2Q_{3n}Q_{3n+1}-Q^2_{3n+1}\nonumber\\
&+&2(Q_{3n}-Q_{3n+1})\sum_{k=0}^{3n-1}Q_k.
\label{pn2}
\end{eqnarray}
Substituting (\ref{pn1}) and (\ref{pn2}) into (\ref{au1}) yields
\begin{eqnarray}
3A_0(t)&=&\sum_{n=0}^{\infty}(Q_{3n}-Q_{3n+1})^2\nonumber\\
&+&\sum_{n=0}^{\infty}(Q^2_{3n-1}+Q^2_{3n}-2Q^2_{3n+1})\nonumber\\
&+&2\sum_{n=0}^{\infty}(2Q_{3n}-Q_{3n-1}-Q_{3n+1})
\sum_{k=0}^{3n-1}Q_k.
\label{aut}
\end{eqnarray}
In the following calculations we use the exact solution (\ref{iqnt}), 
the asymptotic relation 
$I_n(t)\simeq (2\pi t)^{-1/2}\exp(t-n^2/2t)$, 
and the identity\cite{bender}
\begin{equation}
I_{n-1}(t)-I_{n+1}(t)={2n\over t}\,I_n(t).
\label{ident}
\end{equation}
The first sum in the right-hand side of Eq.~(\ref{aut})
behaves as 

\begin{equation}
\sum_{n=0}^\infty (Q_{3n}-Q_{3n+1})^2 \simeq {t^{-3/2}\over 6\sqrt{\pi}}.
\label{negl}
\end{equation}
The second sum in the right-hand side of Eq.~(\ref{aut}) is
undergone by treating $Q^2_{n-1}(t)-2Q^2_{n}(t)+Q^2_{n+1}(t)$ 
as the second derivative, $\partial^2 Q^2_{n}/\partial n^2$, 
which is asymptotically correct.  Using the scaling expression 
(\ref{1-scal}) for $Q_{n}(t)$, this sum is shown to decay as $t^{-2}$ 
in the scaling limit.  Similarly, the computation of the third line in
the right-hand side of Eq.~(\ref{aut}) is simplified by the approximation
$Q_{n-1}(t)-2Q_{n}(t)+Q_{n+1}(t)\simeq \partial^2 Q_{n}/\partial n^2$.
After some algebra, this third term is found to decay as $-(2/3\pi)t^{-1}$
and thus provides a dominant contribution.  The corresponding values 
for $A_1(t)$ and $A_2(t)$ follows from the same kind of computation.
Thus, we finally arrive at the following asymptotic behavior of the
autocorrelation functions:

\begin{equation}
A_0(t)\simeq -{2\over 9\pi t},\quad A_1(t)\simeq {4\over
9\pi t},\quad A_2(t)\simeq -{2\over 9\pi t}.
\label{aut1}
\end{equation}
It is surprising that in the long time limit $A_0$ and $A_2$ exhibit
similar behaviors while the amplitude of the $A_1$ has the opposite sign
and twice bigger.

In the general context of coarsening\cite{bray}, the autocorrelation
function is known to decay as ${\cal L}^{-\lambda}$.  It has been
argued that the exponent $\lambda$ satisfies $d/2\leq
\lambda\leq d$ in $d$ dimensions\cite{fisher}.  Our model
implies $A({\cal L})\sim {\cal L}^{-1}$ ($\lambda=1$) and thus  
coincides with the upper bound as it happens in a few other models, 
{\it e.g.}, in the voter model\cite{elp}.  Most of other 
studies\cite{newman} also found values of the autocorrelation exponent 
satisfying  $d/2\leq \lambda\leq d$ (see, however, Ref.\cite{desai} 
reporting the violation of the upper bound for the conserved dynamics).

\section{THE SPATIAL STRUCTURE}

Turn now to the spatial structure formed as the ballistic annihilation
process proceeds.  Among several quantities characterizing the spatial
distribution we choose the domain size distribution function for which
some analytical results are already available\cite{Piasecki}. 
Let us denote by $\mu_{+-}(x,t)$ the probability density that at 
time $t$ the right nearest neighbor of a $+$
interface is a $-$ interface located at distance $x$ apart.  
Similarly, we introduce $\mu_{++}(x,t)\equiv\mu_{--}(x,t)$ and 
$\mu_{-+}(x,t)$.  The Laplace transform, 
$\hat \mu(z,t)=\int_0^\infty dx\,e^{-xz}\mu(x,t)$,  of these quantities has
been computed exactly\cite{Piasecki}: 
\begin{equation}
\hat \mu_{++}(z,t)={1\over 1+J+2z},
\label{++}
\end{equation}
\begin{equation}
\hat \mu_{-+}(z,t)={S(t)e^{-2zt}\over 1+J+2z},
\label{-+}
\end{equation}
\begin{equation}
\hat \mu_{+-}(z,t)={e^{2zt}\over S(t)}\,{J^2+2z(J-1)\over 1+J+2z},
\label{+-}
\end{equation}
where $S(t)$, the probability for the interface to survive up to time
$t$, is given by Eq.~(\ref{St}), and
\begin{equation}
J\equiv J(z,t)=e^{-2zt}S(t)+
2z\int_0^t d\tau\,e^{-2z\tau}S(\tau).
\label{jzt}
\end{equation}
The solution of Eqs.~(\ref{++})--(\ref{+-})
has been originally derived in an alternative analytical approach
to simpler previous ones\cite{els,Krug}; this approach
of Ref.\cite{Piasecki} has an advantage of being applicable to more 
difficult ballistic annihilation processes like the three velocity 
ballistic annihilation\cite{Droz}.  However, the actual spatial 
characteristics have not been extracted from Eqs.~(\ref{++})--(\ref{+-}).

As a first step, we compute the average length scale 
\begin{equation}
\langle x\rangle={\int_0^\infty dx\,x\mu(x,t)
\over \int_0^\infty dx\,\mu(x,t)}=-{1\over \hat \mu(0,t)}\,
{\partial \hat \mu(z,t)\over \partial z}\bigg|_{z=0}
\label{avescal}
\end{equation}
After straightforward calculations we find the average
size of a domain with boundaries moving in the same direction,
\begin{equation}
\langle x\rangle_{++}={2\over 1+S(t)}\left[\int_0^t d\tau\,S(\tau)
-tS(t)+1\right].
\label{x++}
\end{equation}
Similarly, we find the average domain size in two other situations:
\begin{equation}
\langle x\rangle_{-+}=2t+\langle x\rangle_{++},
\label{x-+}
\end{equation}
and
\begin{eqnarray}
\langle x\rangle_{+-}&=&2{1-S(t)\over S^2(t)}+2t
-{4\over S(t)}\int_0^t d\tau\,S(\tau)\nonumber\\
&+&{2\over 1+S(t)}\left[\int_0^t d\tau\,S(\tau)-tS(t)+1\right].
\label{x+-}
\end{eqnarray}
Making use of the asymptotic relation $S(t)\simeq (\pi t)^{-1/2}$,
we arrive at the following long-time behaviors:
\begin{equation}
\langle x\rangle_{++}\simeq \sqrt{4t\over \pi},\quad
\langle x\rangle_{-+}\simeq 2t,\quad 
\langle x\rangle_{+-}\simeq 2(\pi-3)t.
\label{length}
\end{equation}
The latter result is surprising, one might expect 
that $\langle x\rangle_{+-}$ grows as $\langle x\rangle_{++}$
while in fact it grows much faster.

It is instructive to proceed by computing $\langle x^n\rangle^{1/n}$ 
for arbitrary positive integer index $n$.  One readily expresses 
$\langle x^n\rangle^{1/n}$ via $\hat\mu(z,t)$, {\it e.g.}, 
$\langle x^2\rangle=[\hat \mu(0,t)]^{-1}\,
{\partial^2 \hat \mu(z,t)\over \partial z^2}\big|_{z=0}$.
Any of these quantities can be used to characterize the length scale.  
For domains with dissimilar boundary interfaces one finds 
$\langle x^n\rangle^{1/n}_{+-}\sim \langle x^n\rangle^{1/n}_{-+}
\sim t$, implying that all these distances are characterized by the 
single ballistic length scale, ${\cal L}(t)\sim t$.  In contrast, 
for similar interfaces we get anomalous asymptotic behaviors:
$\langle x^2\rangle^{1/2}\simeq (9\pi)^{-1/4}t^{3/4}$, and generally  
$\langle x^n\rangle^{1/n}_{++}\sim t^{1-1/2n}$ for integer $n$.  
This odd feature indicates the length scale characterizing the average
separation of the nearest similar moving interfaces,  
$\ell(t)=\langle x\rangle_{++}\sim \sqrt{t}$, is just one of the
hierarchy of length scales $\ell_n(t)=\langle x^n\rangle^{1/n}_{++}$.  
All these scales are better
thought as effective scales resulting from the competition between
the two basic scales in the problem, the scale of order one forced by
initial conditions and the ballistic scale of order $t$.

To clarify these results, we compute the inverse Laplace 
transform of Eqs.~(\ref{++},\ref{-+},\ref{+-}). 
We first note that $J(z,t)$ can be rewritten as 
\begin{eqnarray}
J(z,t)&=&2z\int_0^{\infty}d\tau e^{-2z\tau}S(\tau)-\int_t^{\infty}d\tau
         e^{-2z\tau}S'(\tau)\nonumber\\
      &=&-z+\sqrt{z^2+2z}+\int_{2t}^{\infty}d\tau
         e^{-z\tau}{e^{-\tau}I_1(\tau)\over \tau}
\end{eqnarray}
where we have computed the Laplace transform $\hat S(2z)$ and the 
derivative of $S(t)$.  We then expand $\hat\mu_{++}$ to find
\begin{equation}
\hat\mu_{++}(z,t)=\hat a(z)-\hat a(z)^2\tilde b(z,t)
+\hat a(z)^3\hat b(z,t)^2+\ldots
\label{++2}
\end{equation}
where 
\begin{equation}
\hat a(z)=z+1-\sqrt{(z+1)^2-1}
\end{equation}
and
\begin{equation}
\hat b(z,t)=\int_{2t}^{\infty}d\tau
e^{-z\tau}{e^{-\tau}I_1(\tau)\over \tau}.
\end{equation} 
Performing the inverse Laplace transform of $\hat a(z)$ and 
$\hat b(z,t)$, we get
\begin{equation}
a(x)={e^{-x}I_1(x)\over x},\quad b(x,t)={e^{-x}I_1(x)\over
x}\Theta(x-2t).
\label{ab}
\end{equation}
Combining (\ref{++2}) and (\ref{ab}) we finally obtain
\begin{equation}
\mu_{++}(x,t)=a-a^{*2}*b+a^{*3}*b^{*2}-a^{*4}*b^{*3}+\ldots
\label{++3}
\end{equation}
where $f*g=\int_0^x dy\,f(y)g(x-y)$ is the convolution of $f$ and $g$,
$f^{*2}=f*f$, $f^{*3}=f*f*f$,  etc.  

Noting that $a^{*k}(x)=ke^{-x}I_k(x)/x$\cite{laplace}, 
the convolution in the second term of the right-hand side of 
Eq.~(\ref{++3}) can be calculated in the long time limit to yield: 
\begin{eqnarray}
a*a*b&\simeq & {\Theta(\xi)\over \sqrt{2\pi x^3}}\nonumber\\
&\times &\left\{1-e^{-\xi}\left[I_0(\xi)+2I_1(\xi)+I_2(\xi)\right]\right\},
\label{aab}
\end{eqnarray}
where $\xi=x-2t$.  The following terms in Eq.~(\ref{++3}) give 
corrections for $x\geq 4t,6t,\ldots$

The only contribution to $\mu_{++}(x,t)$ for $x<2t$ is the first 
time-independent term $a(x)$. For large $x$, it scales as $x^{-3/2}$. 
According to the mapping of Section II, it is
analogous to the probability that a random walker
starting at the origin first returns back to the origin after $x$ steps. 
A singularity in the second derivative arises at $x=2t$, and weaker and
weaker singularities appear for $x$ being integer multiple of $2t$.
It should be noted that the scale $\ell\sim \sqrt{t}$ does not appear in
this distribution.  The only scales appearing are the scale 
${\cal O}(1)$ characterizing the time-independent contribution $a(x)$, 
and the ballistic scale ${\cal O}(t)$ characterizing next terms.
The asymptotic average distance $\langle x\rangle_{++}\simeq \sqrt{4t/\pi}$ 
of Eq.~(\ref{x++}) is readily obtained by the integration 
$\int_0^{2t}dx\, x\mu_{++}(x,t)=\int_0^{2t}dx\, xa(x)$,
following terms give corrections ${\cal O}(1)$.

Using properties of Laplace transform\cite{laplace}, 
the distributions $\mu_{-+}(x,t)$ and $\mu_{-+}(x,t)$ can be
expressed via $\mu_{++}(x,t)$,
\begin{eqnarray}
\mu_{+-}(x,t)&=&S(t)\,\mu_{++}(x-2t,t)\,\Theta(x-2t)\nonumber\\
\mu_{-+}(x,t)&=&{\mu_{++}(x+2t)\over S(t)},
\end{eqnarray}
thus providing a comprehensive description of the interfaces
distribution in this problem.

\section{SUMMARY AND OUTLOOK}

We have shown that the two-velocity ballistic annihilation
process may be thought as the 3-phase deterministic model of 
coarsening.  This is one of the simplest models of coarsening 
ever known, and we have derived exact solutions
for the generalized first-passage probabilities $P_n(t)$, 
and for the autocorrelation function. 

We have revealed a rich spatial structure arising as the phase 
separation process develops.  In particular, the moments of the 
domain size distribution, $\ell_n(t)=\langle x^n\rangle_{++}^{1/n}$, 
exhibit a variety of scales from the time-independent one 
to the scale linearly growing with time:
\begin{equation}
\ell_n(t)\sim \cases{1& when $n<1/2$,\cr
                     t^{1-1/2n}& when $n>1/2$.\cr}
\label{asymplnt}
\end{equation}
We have argued that only the two extreme scales, the ballistic
one and the scale ${\cal O}(1)$ characterizing the initial distribution,
are important while the others are effective in that they arise as the
result of competition between the extreme scales.  The distribution of
nearest neighbors has shown a non-trivial behavior with 
singularities at each $x$ being an integer multiple of $2t$. 

Using the mapping on a random walk problem introduces in Section II, 
it should be possible to compute the two-point equal-time
correlation function $G(x,t)$ and even the most general
two-point correlation function $C(x,t|0,t')$ which contains both
the equal-time correlation function $G(x,t)\equiv C(x,t|0,t)$ 
and the autocorrelation function $A(t)\equiv C(0,t|0,0)$.  
We were able to solve for $G(x,t)$ for $x\geq 2t$, but the solution is 
very cumbersome so we could not derive clear scaling results.  
Numerical simulations, however, reveal an interesting oscillatory 
behavior of $G(x,t)$.

Another interesting question concerns the extension of the 3-phase
deterministic model to higher dimensions.  
It is very simple to define a 3-color cyclic {\it lattice} model 
in arbitrary dimension\cite{bramson,lpe}.  The problem
is the system does {\it not} exhibit coarsening when $d\geq 2$ and
instead approaches a reactive state with the average number of color
changes growing linearly with time.  However, one can hope that a
proper higher-dimensional extension still exists.

\vskip 0.33in 
We wish to thank S.~Redner for fruitful discussions.  The work of
LF was supported by the Swiss National Foundation, and the work of PLK
was partially supported by ARO (grant DAAH04-93-G-0021) 
and NSF (grant DMR-9219845).

\end{multicols} 
\end{document}